\documentclass[]{mn2e}
\usepackage{psfig,mncite}

\def\figdir{./}

\def\msol{M$_\odot$}

\begin{document}

\title[L dwarf variability]{A mini-survey for variability in early L dwarfs}

\author[F.J. Clarke, B.R. Oppenheimer and C.G. Tinney]{F.J. Clarke$^1$,
B.R. Oppenheimer$^2$ and C.G. Tinney$^3$\\ $^1$Institute of Astronomy,
Madingley Road, Cambridge CB3 0HA, UK.\\ $^2$Department of Astrophysics,
American Museum of Natural History, New York, NY 10024, USA\\
$^3$Anglo-Australian Observatory, PO Box 296, Epping, NSW 2121, Australia\\
fclarke@ast.cam.ac.uk, bro@amnh.org, cgt@aaoepp.aao.gov.au}

\maketitle

\begin{abstract}
We report differential I-band photometry of four early L-dwarfs obtained to
study variability. We detect variability on the timescale of hours in two
objects, 2M0746425+200032 (at a level of 0.007\,mag -- 6.5 sigma) and
2M1108307+683017 (0.012\,mag -- 5 sigma). We also place upper limits of
0.02\,mag (1 sigma) on the variability of two others.
\end{abstract}


\section{Introduction}

The recently catagorised L-dwarfs \cite{kirkpatrick99} have rapidly become
an important part of stellar astrophysics. L-dwarfs lie below M-dwarfs in
terms of their temperatures, and their spectra are dominated by molecules
such as TiO, FeH and CrH. Dust formation also plays a very significant
rol\'e in the atmospheric physics of early L-dwarfs (by L5-7, the major
dust species have begun to settle below the photosphere). The physics of
dust formation is a complicated subject, but one which is critical to our
understanding of L dwarfs. Variability studies give us a tool to probe the
sub-global differences in the composition of the atmosphere. Other
processes, such as the flaring seen in M dwarfs, may also contribute to
variability in L dwarfs.

In this paper we report time-series photometry of four field L dwarfs. Two
objects, 2MASS1108307+683017 and 2MASS0746425+200032 are found to be
variable at the level of 0.012\,mag (5$\sigma$) and 0.007\,mag (6.5$\sigma$)
respectively on the timescale of hours. We did not detect any variability
above the noise level of $\sim$0.02\,mag (1$\sigma$) from the other two
targets (2MASS1146345+223053 and DENIS0909571-065806).

\section{Observations and Reduction}

Data were obtained with PFCam on the 3-m Shane telescope at Lick
Observatory on the nights of 2002 January 21 and 23 UT. Conditions were
photometric on the first night, but patchy cloud interupted observations on
the 23rd. PFCam has a 2048$\times$2048 CCD array with a pixel scale
of 0.296\arcsec/pixel. The full field of view,
10.1\arcmin$\times$10.1\arcmin, is more than sufficent for our purpose
(high cadence, high precision differential photometry). We therefore used
only a small section of the chip to improve read-out times. Chip windows
were choosen to select the target, and 5 or 6 comparison stars of similar
magnitude. The chip was also binned where possible to further reduce
overheads, such that we collected images at 2 minute intervals on
average. All observations were made through the I-band filter where the
photon count from the extremely red L dwarfs is maximised. Calibration
frames, consisting of bias frames, dome and sky flats were obtained for all
chip configurations. A log of observations, including window sizes and chip
binning, is given in table~\ref{tab:observations}.

Data were reduced with IRAF\footnote{The Image Reduction and Analysis
Facility (IRAF) is distributed by NOAO, which is operated by AURA, Inc.,
under contract to the NSF.}. Ten bias frames were median combined, and this
was then subtracted from all data and flat field frames. An average
normalised flat field was then constructed from sky flats. We divided the
science data by the appropriate flat field. This provided images flattened
to the 0.25\% level. Fringing is visible at the $\sim$1\% level on the
scale of $\sim$10\arcsec. We could not remove this, as we did not obtain
enough pointings to build a fringe map.

During the first night's observing, we discovered PFCam's shutter was
working incorrectly, with only one blade of the traveling shutter system
functioning. As the travel time of the blade is on the order of the
milliseconds, a part in 10$^6$ for most science frames, this contributes
neglibly to the photometric errors. The effect is however apparent in short
(0.5-4s) dome flats. For these frames, we fitted the gradient with a linear
function, and subtracted the fit.

In some science frames, the peak counts in a few pixels approached the
full-well depth of the CCD (65536 ADU). We obtained data to test the
linearity of the chip, and found it to be linear to $\sim$60000 ADU. Any
frames with peak counts above this level were removed from the analysis
below.

Photometry was performed with the APPHOT package within IRAF. We determined
the FWHM for each frame within an observing sequence, and picked an
aperture $\sim$1.5$\times$ larger than the average FWHM for that
sequence. This aperture was used on all frames within that
sequence. Magnitudes and errors were calculated from the fluxes given by
APPHOT in the manner described by \scite{everett01}, including an estimate
of informal errors from fringing and imperfect flatfield correction. An
ensemble of bright stars were combined to give a zero-point magnitude for
each frame (m$_{\textrm{\tiny ens}}$), which was then subtracted from the
measured magnitude of the target (m$_{\textrm{\tiny tar}}$) to build a
differential lightcurve. Hence, an increase in $\delta$m (m$_{\textrm{\tiny
tar}}$-m$_{\textrm{\tiny ens}}$) represents a dimming of the object. A
selection of comparison stars (not included in the ensemble stars) were
also analysed to ensure any variability detected is intrinsic to the
target. In no case did any of the ensemble stars display any variability.

\begin{table*}
\begin{tabular}{cccccccccc}
Object$^1$ & Date & Start UT & End UT & \# Frames & Exp time &
\multicolumn{2}{c}{CCD configuation} \\ & 2002 Jan & & & & (s) & Window
size & binning\\
\hline
2M0746425+200032 & 21 & 02:48 & 03:33 & 28 & 45 & 5.1\arcmin$\times$5.1\arcmin & 2$\times$2 \\
2M0746425+200032 & 21 & 03:33 & 04:55 & 51 & 45-30 & 3.7\arcmin$\times$3.7\arcmin & 2$\times$2 \\
2M0746425+200032 & 21 & 04:55 & 09:22 & 121 & 120-60 & 2.2\arcmin$\times$2.2\arcmin & 1$\times$1 \\
2M1108307+683017 & 21 & 09:52 & 14:23 & 119 & 100 & 5.1\arcmin$\times$5.1\arcmin & 2$\times$2 \\
D0909571-065806 & 23 & 07:51 & 08:20 & 6 & 300 & 2.2\arcmin$\times$2.2\arcmin & 2$\times$2 \\ 
D0909571-065806 & 23 & 09:53 & 11:57 & 26 & 300-200 & 2.2\arcmin$\times$2.2\arcmin & 2$\times$2 \\ 
2M1146345+223053 & 23 & 12:11 & 14:20 & 22 & 300 & 2.2\arcmin$\times$2.2\arcmin & 2$\times$2 \\ 
\end{tabular}
\caption{Log of observations obtained with PFCam on the 3-m Shane Telescope
at Lick Observatory. Object names starting with 2M come from the 2MASS
survey \protect\cite{kirkpatrick99}. D0909-0658 was discovered by the DENIS
survey \protect\cite{delfosse97}.}
\label{tab:observations}
\end{table*}

\section{Results}

\begin{figure*}
\begin{tabular}{cc}
\psfig{file=\figdir/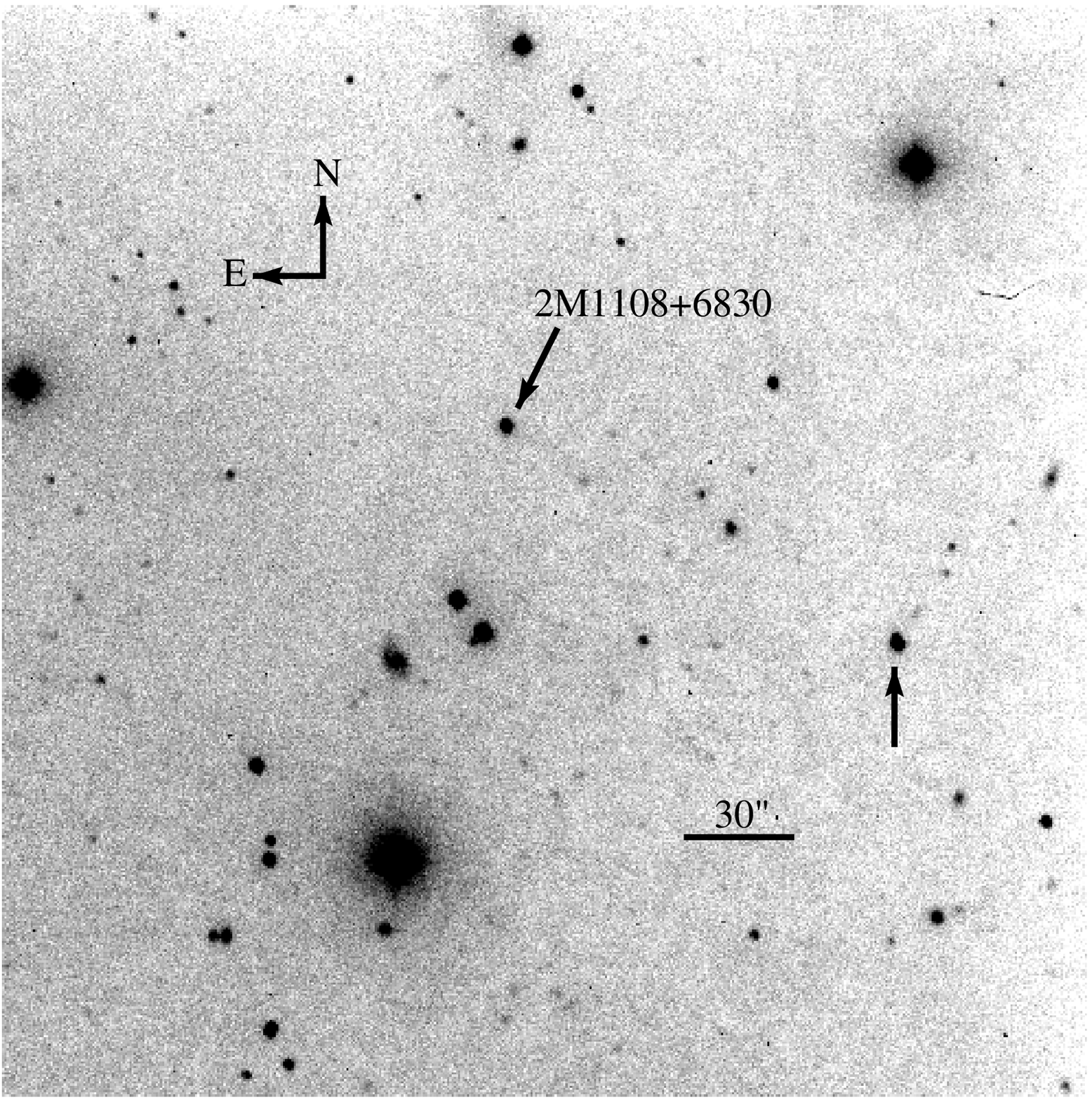,width=8.cm,height=8.cm}&
\psfig{file=\figdir/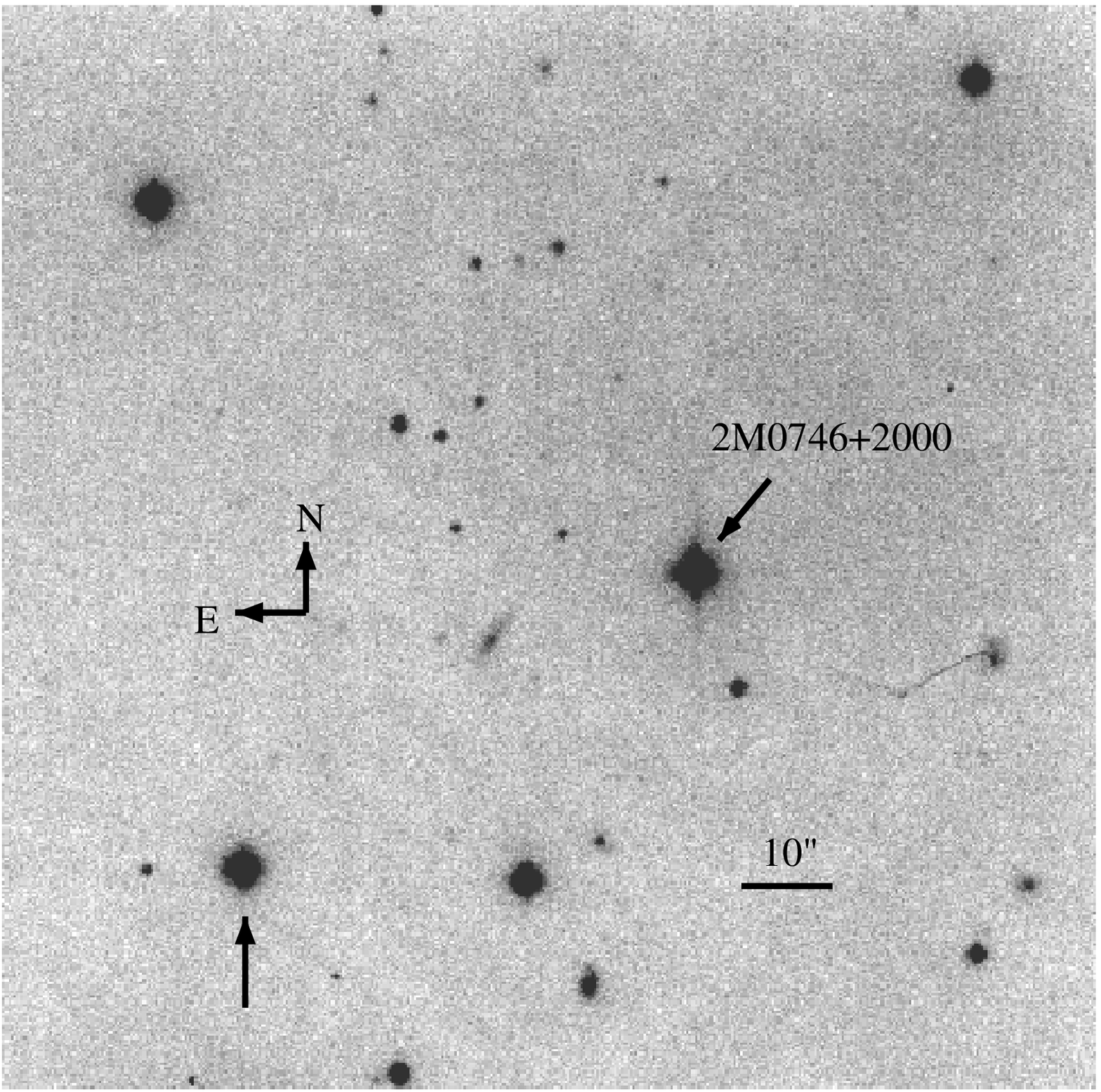,width=8.cm,height=8.cm}\\ 
\psfig{file=\figdir/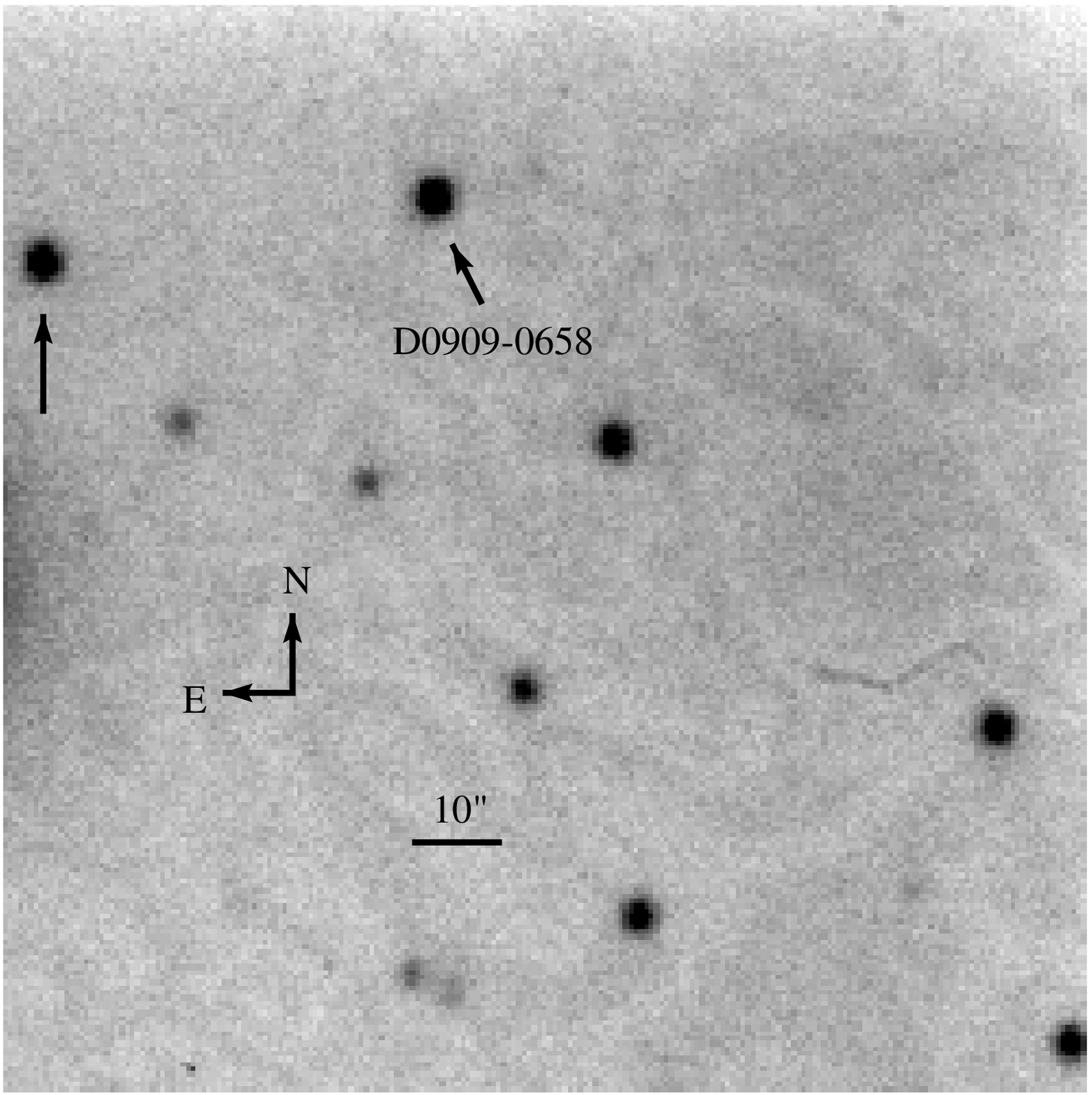,width=8.cm,height=8.cm} & 
\psfig{file=\figdir/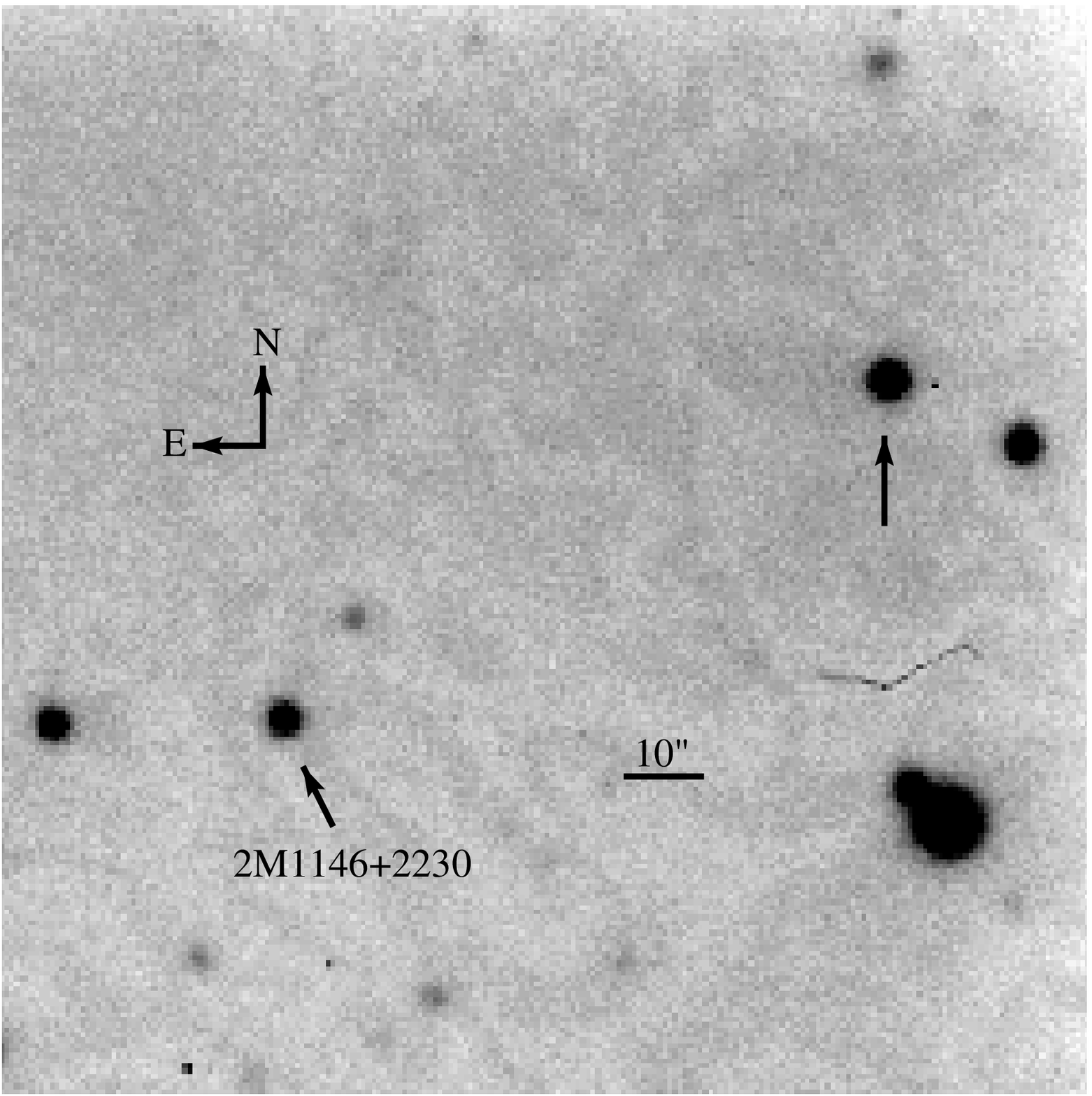,width=8.cm,height=8.cm} \\
\end{tabular}
\caption{Finder chart for all four targets, indicating the position of the
target and the comparison stars (marked with arrows). Note the difference
in scale of the 2M1108+6830 field.}
\label{fig:finders}
\end{figure*}

\subsection{2MASSJ1108307+683017}

\begin{figure}
\psfig{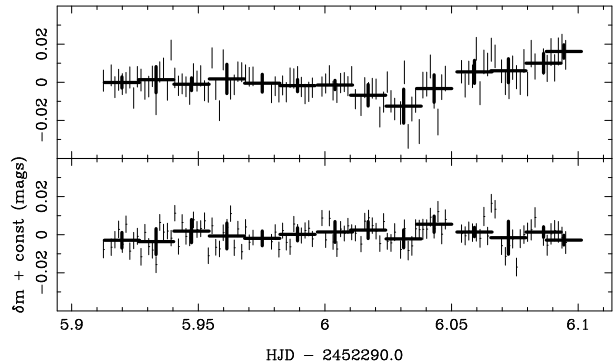}
\caption{I band differential lightcurve of 2M1108307+683017 (upper panel)
and the comparison star indicated in Figure~\ref{fig:finders} (lower
panel). Individual measurements are shown with narrow lines, with the
lightcurve averaged into 20 minute bins shown in thick crosses. 2M1108
shows a dimming (increasing $\delta$m) of approximately 0.012\,mag over
$\sim$2 hrs. The comparison star stays constant to $\pm$0.003\,mag
throughout the night. Both lightcurves have been mean subtracted.}
\label{fig:2M1108lc}
\end{figure}

2M1108+6830 is an L1 dwarf. \scite{gizis00} report an H-$\alpha$ equivalent
width of 7.8\AA, which they suggest indicates it is possibly an older
member of the stellar population, rather than a youthful brown dwarf. We
observed 2M1108 for 4.5 hours with a sampling rate of 2 minutes (100s
integrations). To obtain sufficent ensemble stars (at least 3), we used a
field of 5.1\arcmin$\times$5.1\arcmin\ with 2x2 pixel binning
(Figure~\ref{fig:finders}). Figure~\ref{fig:2M1108lc} shows the resulting I
band lightcurves for 2M1108 and a comparison star (which was not part of
ensemble stars). It is clear from this figure that 2M1108 becomes
$\sim$0.012\,mag fainter during the last 1.5 hours of observation, whilst
the comparison star stays constant to within $\pm$0.003\,mag throughout the
observation. The standard deviation of the binned comparison lightcurve is
0.0025\,mag. Our detection of variability in the binned lightcurve is
therefore significant at the 5$\sigma$ level. There may be a slight
brightening in 2M1108 just before it fades (AJD\footnote{AJD =
HJD-2452290.0}=6.03), but this is detected only at the 2$\sigma$ level due
to an increase in noise in this region.

Although the comparison star shows no variability, we must consider the
possibility that 2M1108's variability is caused by second order colour
effects (see \scite{young91} for a full discussion). The L dwarfs have much
redder spectral energy distributions than the ensemble stars, and hence
have a longer {\em effective} wavelength in the I filter. This means the L
dwarfs have a smaller effective extinction coefficent, which would manifest
itself as an apparent brightening in the differential lightcurve with
increased airmass. However, the data exhibit the opposite behaviour, since
the observation started at an airmass of 1.15 and ended at 1.3. The
detected dimming of 2M1108 we see is therefore unlikely to be caused by
airmass effects.

The H-$\alpha$ emission from 2M1108 implies chromospheric
activity. However, it is not clear if the variability we detect is related
to this. The variability may alternatively be due to photospheric features
(magnetic spots or dust clouds) which modulate the brightness as the object
rotates. There is no periodic signal in our data, but our temporal sampling
clearly does not cover a sufficent range to exclude this possibility. Based
on $v \sin i$ measurements \cite{basri00}, we expect the typical rotation
period of an L dwarf to be on the order of 6-12 hours.

\subsection{2MASSJ0746425+200032}

2M0746+2000 is infact a close (0.22\arcsec) binary of roughly equal mass
\cite{reid01}. We do not resolve the binary in our observations, and
therefore refer to the whole system as 2M0746 rather an either individual
component. We discuss how this may effect our results later.

2M0746 has a spectral type of L0.5. \scite{schweitzer01} detect no Li in
its spectra, and therefore estimate a minimum mass of $\sim$0.06\msol for
each component. Due partly to its binary nature, 2M0746 is one of the
brightest known L dwarf systems, having an I band magnitude of 15.2. It
also lies in a rich stellar field, making it an ideal target for accurate
differential photometry. We observed this target for nearly 6.5 hours with
a sampling rate of 1-2 minutes. Changes in seeing required us to adapt our
observing strategy (to avoid saturating the target). The majority of the
usable images measure 2.2\arcmin$\times$2.2\arcmin, and contain the target
and 4 bright comparison stars (Figure~\ref{fig:finders}). These stars are
also visible in all previous configurations used on this object.

\begin{figure}
\psfig{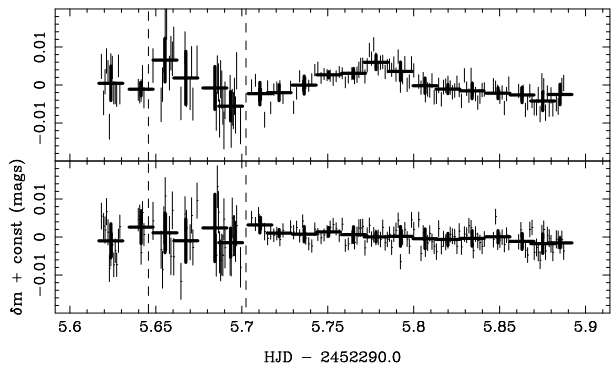}
\caption{I band lightcurve of 2M0746425+200032 (upper panel) and an
apparently invariant comparison star (lower panel). Individual measurements
are shown as narrow lines, with the lightcurve averaged into 20 minute bins
shown with thick crosses. 2M0746 clearly changes in brightness (dims) by
$\sim$0.007\,mag on the timescale of 3 hours while the comparison star
stays constant to $\sim$0.002\,mag. The dashed vertical lines at times 5.64
and 5.70 indicate the changes in chip configuration
(Table~\ref{tab:observations}). Each configuration has a different
zeropoint, and all three have been individually mean
subtracted. Differences between the configurations are therefore not
significant.}
\label{fig:2M0746lc}
\end{figure}

The derived I band lightcurves are shown in Figure~\ref{fig:2M0746lc}.
2M0746 shows variability in two of the three configurations. In the second
configuration (AJD=5.64-5.70), 2M0746 brightens ($\delta$m decreases) by
$\sim$0.01\,mag while the comparison stays constant to within
0.005\,mag. However, due to the increased noise in this configuration, this
is only signficant at the 2$\sigma$ level. During the third configuration
(AJD=5.70-5.89), 2M0746 fades by $\sim$0.007\,mag around AJD=5.78 before
returning to its original brightness. The comparison object is constant to
$\sim$0.002\,mag, and its binned lightcurve has a standard deviation of
0.0011\,mag. Our detection of variability in the third configuration is
therefore significant at the 6.5$\sigma$ level.

It is possible that the brightening we see from 5.66 to 5.7 is analogous to
the brightening from 5.78 to 5.82. This may suggest that we are seeing
rotational modulation of the lightcurve on a period of $\sim$3 hours. If
this were the case, we would expect to see the next minima at
AJD$\simeq$5.9 - just after the end of our observations. Three hours would
be a fast, but not unprecedented rotational period for a brown dwarf. For
example, Kelu-1 has a rotational period of 1.8 hours
\cite{clarke02}.

Unfortunately, the binary nature of 2M0746 complicates matters
significantly. As we are observing an unresolved binary, we are essentially
sampling the combined lightcurve of two objects. It is not therefore
possible to attribute specific events in the lightcurve to either
object. In addition, the binarity will tend to ``wash-out'' variability
from either component of the binary; for example, 1\% variability from one
{\em component} will result in only 0.5\% measured variability from the
{\em system}.

2M0746 is clearly an interesting target for future variability
observations, especially if the binary can be resolved.

\subsection{DENIS0909571-065806}

\begin{figure}
\psfig{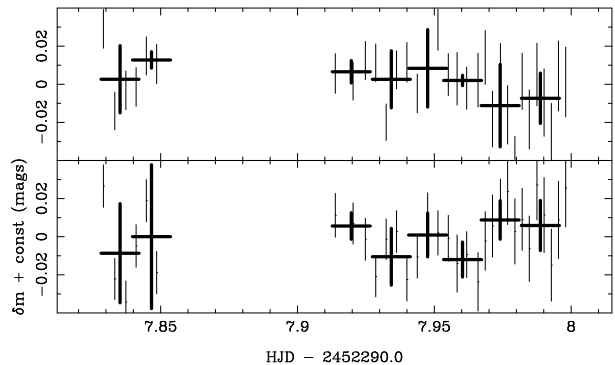}
\caption{Lightcurve of D0909571-065806 (upper panel) and the comparison
star indicated in Figure~\ref{fig:finders} (lower panel). The is no
significant variability from the target, with the scatter in both
lightcurves at the level of $\sigma_i$=0.02\,mag.}
\label{fig:D0909phot}
\end{figure}

\scite{martin99} classify D0909-0658 as an L0
dwarf. Figure~\ref{fig:finders} identifies it along with the comparison
star whose lightcurve is plotted in Figure~\ref{fig:D0909phot}. Seeing on
this night was $\sim$2 arcsec poorer than the first night, requiring longer
(300s) exposures to obtain the necessary signal to noise ratio. D0909-0658
was observed in two batches, seperated by $\sim$90
minutes. Figure~\ref{fig:D0909phot} shows no significant indication of
variability either between or within the two sequences. We therefore place
an upper limit on the variability of D0909-0658 of
$\sigma_{\textrm{i}}<$0.02\,mag (1 sigma) on the timescale of 0.5-4 hours.

\subsection{2MASS1146345+223053}

2M1146+2230 is actually a close equal brightness binary with a separation
of 0.3\arcsec\ \cite{koerner99}. \scite{schweitzer01} have detected Li in
the combined spectrum of 2M1146, confirming its status as a brown dwarf
binary with each component having M$<$0.06\msol. In addition, an earlier
type background star is located 1\arcsec\ away from the binary
\cite{kirkpatrick99}. The seeing during our observations (3 arcsec) did not
allow us to resolve the binary or to detect the background star.

\begin{figure}
\psfig{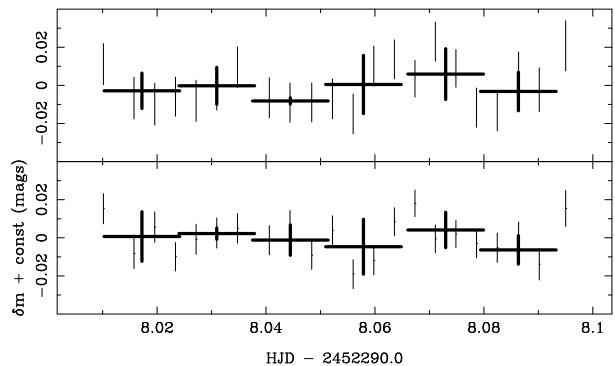}
\caption{Lightcurve of 2M1146345+223053 (upper panel) and the comparison
star (lower panel) denoted in figure~\ref{fig:finders}. Both lightcurves
are consistent with non-variability within the errors. This gives an upper
limit of $\sigma_i<$0.02\,mag for the variability of 2M1146 over $\sim$2
hours.}
\label{fig:2M1146lc}
\end{figure}

\scite{bailerjones01} have previously made a marginal detection of
variability in this object at the level of $\sim$0.015 mag. Poor seeing did
not allow us to reach this level of accuracy. The lightcurve
(Figure~\ref{fig:2M1146lc}) shows 2M1146 is constistent with being
non-variable at the noise level of $\sim$0.02\,mag (1 sigma) during our
observations.

\section{Conclusions}

We have detected variability in time series I band observations of two
field L dwarfs (2M0746425+200032 and 2M1108307+683017), and placed upper
limits on the variability of two more (2M1146345+223053 and
D0909571-065806). The cause of variability is unclear, but it may be due to
imhomogenous structures within the photosphere. In the case of 2M0746+2000,
weak evidence exists for a periodicity of $\sim$3 hrs. This could be
explained by a photospheric feature and a 3 hour rotation period.

In now seems clear that a significant fraction of, if not all, L-dwarfs are
variable at the 1-2\% level, and that very few have much larger variability
(\pcite{clarke02}, \pcite{bailerjones01}, \pcite{martin01}). In this study,
both stars for which we obtained photometry better than 1\% exhibited
variability. Future surveys for variability in L-dwarfs require a
photometric precision better than 0.5\%.

\section{Acknowledgements}

This research is based on data obtained with the Shane telescope at Lick
Observatory, California.  We would like to thank the staff of Lick
Observatory for their help and hospitality during our run, especially the
coffee. We also thank Simon Hodgkin for useful discussions and comments,
and also the referee, John Gizis, for his helpful comments. FJC
acknowledges the support of a PPARC studentship award during the course of
this research. BRO is supported by a Hubble fellowship, STScI grant
HST-HT-01122.02-A.

\bibliography{lickvar}
\bibliographystyle{mn}

\end{document}